\documentclass[twocolumn,aps,prl,showpacs,superscriptaddress,groupedaddress]{revtex4}
\usepackage[latin9]{inputenc}
\usepackage{color}
\usepackage{amsmath}
\usepackage{amssymb}
\usepackage{graphicx}
\usepackage{esint}
\usepackage[unicode=true,pdfusetitle,
 bookmarks=true,bookmarksnumbered=false,bookmarksopen=false,
 breaklinks=true,pdfborder={0 0 1},backref=false,colorlinks=true]
 {hyperref}
\hypersetup{
 linkcolor=red,urlcolor=blue,citecolor=blue,anchorcolor=blue}
\usepackage{breakurl}

\makeatletter
\@ifundefined{textcolor}{}
{%
 \definecolor{BLACK}{gray}{0}
 \definecolor{WHITE}{gray}{1}
 \definecolor{RED}{rgb}{1,0,0}
 \definecolor{GREEN}{rgb}{0,1,0}
 \definecolor{BLUE}{rgb}{0,0,1}
 \definecolor{CYAN}{cmyk}{1,0,0,0}
 \definecolor{MAGENTA}{cmyk}{0,1,0,0}
 \definecolor{YELLOW}{cmyk}{0,0,1,0}
}

\usepackage{dcolumn}\usepackage{bm}

\makeatother

\begin{document}

\title{A distinguishable single excited-impurity in a Bose-Einstein condensate}

\author{Javed Akram}

\email{javedakram@daad-alumni.de}

\affiliation{Department of Physics, COMSATS, Institute of Information Technology
Islamabad, Pakistan}

\affiliation{Institute f\"{u}r Theoretische Physik, Freie Universit\"{a}t Berlin, Arnimallee
14, 14195 Berlin, Germany}

\date{\today}
\begin{abstract}
We investigate the properties of a distinguishable single excited state impurity pinned
in the center of a trapped Bose-Einstein condensate (BEC) in a one-dimensional
harmonic trapping potential by changing the bare mass of the impurity
and its interspecies interaction strength with the BEC. We model our
system by using two coupled differential equations for the condensate
and the single excited-impurity wave function, which we solve numerically.
For equilibrium, we obtain that an excited-impurity induces two bumps
or dips on the condensate for the attractive- or repulsive-interspecies
coupling strengths, respectively. Afterwards, we show that the excited-impurity
induced imprint upon the condensate wave function remains present
during a time-of-flight (TOF) expansion after having switched off
the harmonic confinement. We also investigate shock-waves or gray-solitons by switching off the interspecies coupling strength
in the presence of harmonic trapping potential. During this process,
we found out that the generation of gray bi-soliton or gray quad-solitons (four-solitons)  
depends on the bare mass of the excited-impurity in a harmonic trap. 
\end{abstract} 


\maketitle

\section{Introduction} 

Certainly, the physics of trapped condensates has emerged as one of
the most exciting fields of physics in last few decades. During last few
years, substantial experimental and theoretical progress has been
made in the study of the properties of this new state of matter. The remarkable experimental realization of a Bose-Einstein condensate
mixture composed of two spin states of $^{87}\text{Rb}$ \cite{Andrews97,PhysRevLett.78.586}
has a rapid compelling interest in the physics of a new class of quantum
fluids: the two or more species Bose-Einstein condensates 
\cite{PhysRevLett.81.1539,PhysRevLett.89.060403,PhysRevLett.101.040402,PhysRevA.84.011603}. 
Multi-species condensates (MSC) offer new degrees of freedom, which
give rise to a rich set of new issues \cite{RevModPhys.83.1405,PhysRevA.92.023630}, at the heart of many of these issues is the presence
of interspecies interactions and the resulting coupling of the two
condensates. Previous theoretical treatments have shown that due to
interspecies interactions, the ground state density distribution of
MSC can display novel structures that do not exist in a one-species
condensate \cite{PhysRevLett.77.3276,PhysRevLett.80.1130}. The investigation
of a hybrid system requires progress on several different fronts.
For example,  from last few decades, many theoretical and experimental researchers
focus on a single-particle impurity control in a many-body system
for the detection and engineering of strongly correlated quantum states
\cite{Bakr09,Sherson10,Serwane11,Ratschbacher12,PhysRevA.90.033601,Schurer15}.

\begin{figure*}
\includegraphics[width=16cm,height=6.5cm]{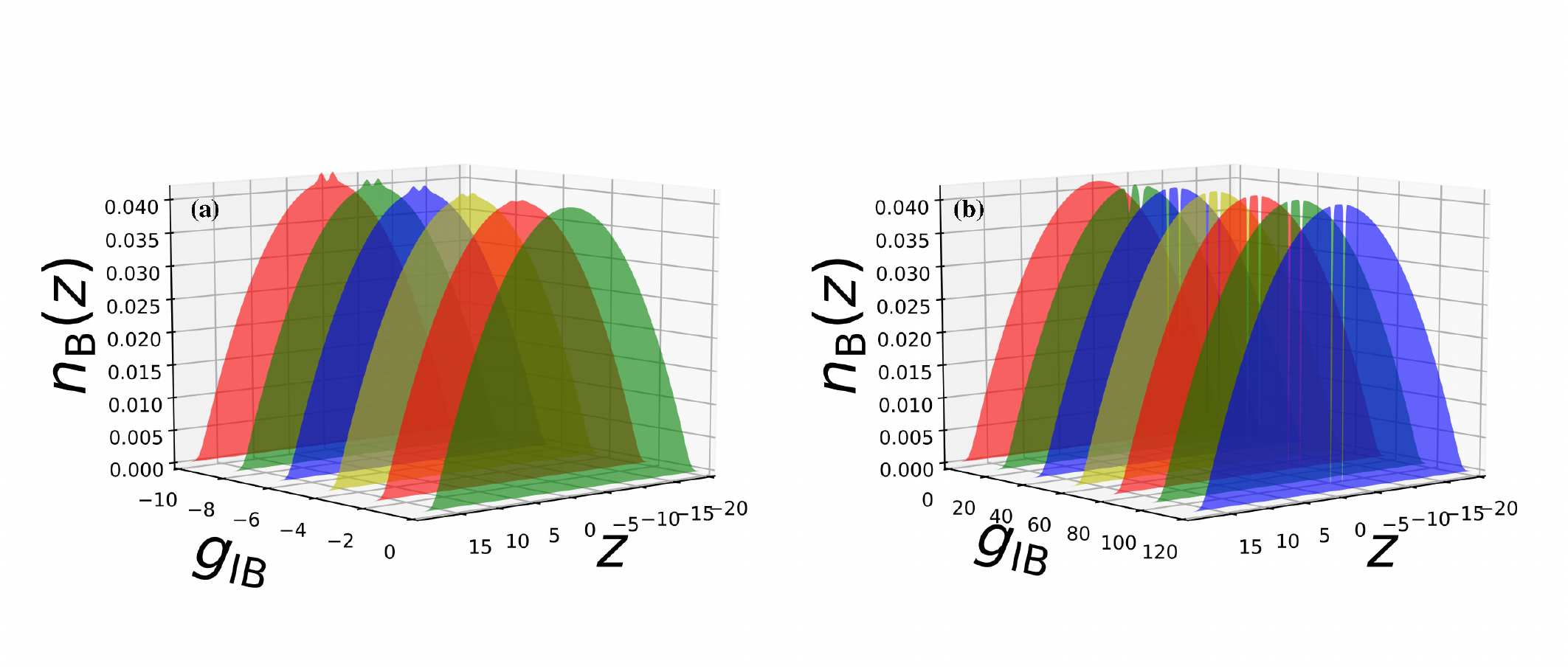}

\vspace*{-0.9cm}
 \protect\protect\caption{Numerical density profile of BEC for (a) negative and
(b) positive values of $g_{\text{IB}}$ as mentioned in the figures
for the experimental coupling strength value $G_{\text{B}}=4.71$
in dimensionless units. \label{Fig1}}
\end{figure*}

Individually controllable impurities in a quantum gas grants access
to a huge number of proposed novel applications \cite{PhysRevLett.102.230402,PhysRevLett.103.150601,Zipkes10,PhysRevLett.105.133202}.
In the direction of quantum information processing, atomtronics applications
are envisioned with single atoms acting as switches for a macroscopic
system in an atomtronics circuit \cite{PhysRevLett.93.140408}; two
impurity atoms immersed in a quantum gas can be employed for a transfer
of quantum information between the atoms \cite{PhysRevA.71.033605},
or individual qubits can be cooled preserving the internal state of
coherence \cite{PhysRevA.69.022306,PhysRevLett.97.220403}. Adding
impurities and, hence, polarons one by one allow experimentalists
to track the transition even to the many-body regime and, moreover,
yield information about spatial cluster formation \cite{Klein07,PhysRevA.84.063612,Santamore11,Widera15}.
Furthermore, adding single impurities one-by-one to an initially integrable
system, such as a quasi one-dimensional Bose gas \cite{kinoshita06},
allows one to controllably induce the thermalization of a non equilibrium
quantum state. The coupling of impurities with condensed matter helps
to understand material properties such as molecule formation and electrical
conductivity \cite{PhysRevA.78.023610,Lercher11,Spethmann012,PhysRevLett.109.235301,Balewski13}.

Recently, we investigated static and dynamical propertied of 
a single ground state impurity $^{133}\text{Cs}$ in the center of a trapped $^{87}\text{Rb}$
BEC \cite{PhysRevA.93.033610}. We studied the physical similarities and differences of bright shock
waves and gray/dark bi-solitons, which emerge for an
initial negative and positive interspecies coupling constant, respectively \cite{PhysRevA.93.033610}. 
In this letter, we want to extend our previous 
work to the single excited impurity in the center of a trapped BEC.
In the following, in Sec.~\ref{PS2}, we define two coupled one-dimensional
differential equations (1DDEs), where one equation is nothing but
a quasi one dimensional Gross-Pitaevskii equation (1DGPE) with a potential
term stemming from the excited-impurity and the second equation is
a typical Schr\"{o}dinger wave equation with an additional potential originating
from the BEC. Afterwards, we show that the single excited state impurity
(SESI) imprint upon the condensate wave function strongly depends
upon whether the effective SESI-BEC coupling strength is attractive
or repulsive. Subsequently, the dynamics of the SESI imprint upon
the condensate wave function is discussed in detail in Sec.~\ref{PS3}.
Here, we note that for an attractive interspecies coupling strength
the excited-impurity imprint does not decay but decreases for a repulsive
interspecies coupling strength in a time-of-flight (TOF). In the same
Sec.~\ref{PS3}, we discuss the creation of shock-waves or gray quad-solitons
in a harmonic trap, by switching off the attractive or repulsive Rb-Cs
coupling strength. Finally, in Sec.~\ref{PS6} we make concluding
remarks and comment on the realization of the proposed model system.

\section{Model}

\label{PS2}We assume an effective quasi one-dimensional setting with
$\omega_{\text{z}}\ll\omega_{\text{r}}$, so the theoretical model
for describing the time evolution of two-component BECs is the following
coupled GP equations as 
\begin{align}
i\hbar\frac{\partial}{\partial t}\psi(z,t)=\left\{ -\frac{\hbar^{2}}{2m_{B}}\frac{\partial^{2}}{\partial z^{2}}
+\frac{m_{\text{B}}\omega_{\text{z}}^{2}}{2}z^{2}+G_{\text{IB}} \lvert \psi_{\text{I}}\left(z,t\right)\lvert^{2}\right.\nonumber \\
\left.+G_{\text{B}}\lvert\psi(z,t)\lvert^{2}\right\} \psi(z,t),\label{Eq1}\\
i\hbar\frac{\partial}{\partial t}\psi_{\text{I}}\left(z,t\right)=\left\{ -\frac{\hbar^{2}}{2m_{\text{I}}}\frac{\partial^{2}}
{\partial z^{2}}+\frac{m_{\text{I}}\omega_{\text{Iz}}^{2}}{2}z^{2} +G_{\text{BI}}\lvert\psi\left(z,t\right)\lvert^{2}\right\}  \nonumber \\
 \times \psi_{\text{I}}\left(z,t\right).\label{Eq2}
\end{align}
where $\psi(z,t)$ denotes the macroscopic condensate wave function
for the $^{87}\text{Rb}$ BEC and $\psi_{\text{I}}\left(z,t\right)$
describes $^{133}\text{Cs}$ single excited state impurity with $z$
being the spatial coordinate, here $m_{\text{B}}$ and $m_{\text{I}}$
stands for the mass of the $^{87}\text{Rb}$ and $^{133}\text{Cs}$
atom, respectively. In the above Eq.~(\ref{Eq1}), $G_{\text{B}}=2N_{\text{B}}a_{\text{B}}\hbar\omega_{\text{r}}$
represents the one-dimensional $^{87}\text{Rb}$ coupling strength,
where $N_{\text{B}}=200$ denotes the number of $^{87}\text{Rb}$
atoms, and the s-wave scattering length is $a_{\text{B}}=94.7~{\rm {a}_{0}}$
with the Bohr radius ${\rm {a}_{0}}$. In the first equation, $G_{\text{IB}}=N_{\text{I}}g_{\text{IB}}$
stands for the impurity-BEC coupling where $g_{\text{IB}}=2a_{\text{IB}}\hbar\omega_{\text{r}}f\left(\omega_{\text{Ir}}/\omega_{\text{r}}\right)$
and $f\left(\omega_{\text{Ir}}/\omega_{\text{r}}\right)=\left[1+\left(m_{\text{B}}/m_{\text{I}}\right)
\right]/\left[1+\left(m_{\text{B}}\omega_{\text{r}}\right)/\left(m_{\text{I}}\omega_{\text{Ir}}\right)\right]$
represents a geometric function \cite{PhysRevA.93.033610}, which
depends on the ratio of the trap frequencies, and $N_{\text{I}}=1$
stands for the number of excited-impurity atoms, and $a_{\text{IB}}=650~{\rm {a}_{0}}$
expresses the effective Rb-Cs s-wave scattering, which can be modified by
Feshbach resonance \cite{Lercher11,PhysRevA.79.042718,PhysRevA.85.032506,PhysRevA.84.013618,PhysRevA.90.013633}.
Here $G_{\text{BI}}=N_{\text{B}}g_{\text{IB}}$
describes the BEC-impurity coupling strength. Presently, we let that
the excited-impurity and the BEC are in the same trap, therefore,
$\omega_{\text{Ir}}=\omega_{\text{r}}=2\pi\times0.179~\text{kHz}$
and $\omega_{\text{Iz}}=\omega_{\text{z}}=2\pi\times0.050~\text{kHz}$.
When the impurity atom decays to its ground state, it emits photon
with energy corresponding to the difference between the excited and
ground states of $^{133}\text{Cs}$ atom. In our case, we let that
the decay of the excited-impurity atom is damped by using the quantum
zeno effect \cite{PhysRevA.41.2295,PhysRevLett.87.040402,RevModPhys.75.281}. 
The quantum zeno effect is an aspect of quantum mechanics, where a particle's wave function 
time evolution can be seized by measuring it frequently enough with respect to some 
chosen measurement setting. If the period between measurements is short enough, 
the wave function usually collapses back to the initial state \cite{PhysRevA.41.2295,PhysRevLett.87.040402,RevModPhys.75.281}.
In order to make Eq.~(\ref{Eq1}) and Eq.~(\ref{Eq2}) dimensionless,
we establish the dimensionless coordinate as $\tilde{z}=z/l_{\text{z}}$,
the dimensionless time as $\tilde{t}=\omega_{\text{z}}t$, and the
dimensionless wave function as $\tilde{\psi}=\psi\sqrt{l_{\text{z}}}$($\tilde{\psi}_{\text{I}}=\psi_{\text{I}}\sqrt{l_{\text{z}}}$),
where the oscillator length $l_{\text{z}}=\sqrt{\hbar/(m_{\text{B}}\omega_{\text{z}})}$
is given by 28742.3 ${\rm {a}_{0}}$ for the above mentioned experimental
values. With this Eq.~(\ref{Eq1}) and Eq.~(\ref{Eq2}) can be rewritten
in dimensionless form
\begin{align}
i\frac{\partial}{\partial\tilde{t}}\tilde{\psi}\left(\tilde{z},\tilde{t}\right)
=\left\{ -\frac{1}{2}\frac{\partial^{2}}{\partial\tilde{z}^{2}}+\frac{\tilde{z}^{2}}{2}
+\tilde{G}_{\text{B}}\lvert\tilde{\psi}\left(\tilde{z},\tilde{t}\right)\lvert^{2}+\tilde{G}_{\text{IB}}\right.\nonumber \\
\left.\times\lvert\tilde{\psi}_{\text{I}}(\tilde{z},\tilde{t})\lvert^{2}\right\} \tilde{\psi}\left(\tilde{z},\tilde{t}\right),\label{Eq03}\\
i\frac{\partial}{\partial\tilde{t}}\tilde{\psi}_{\text{I}}\left(\tilde{z},\tilde{t}\right)=\left\{ -\frac{\tilde{\alpha}^{2}}{2}\frac{\partial^{2}}{\partial z^{2}}+\frac{\tilde{z}^{2}}{2\tilde{\alpha}^{2}}+\tilde{G}_{\text{BI}}\lvert\tilde{\psi}(\tilde{z},\tilde{t})\lvert^{2}\right\} \nonumber \\
\times\tilde{\psi}_{\text{I}}\left(\tilde{z},\tilde{t}\right).\label{Eq4}
\end{align}

\begin{figure*}
\begin{center}
 \includegraphics[width=16cm,height=7cm]{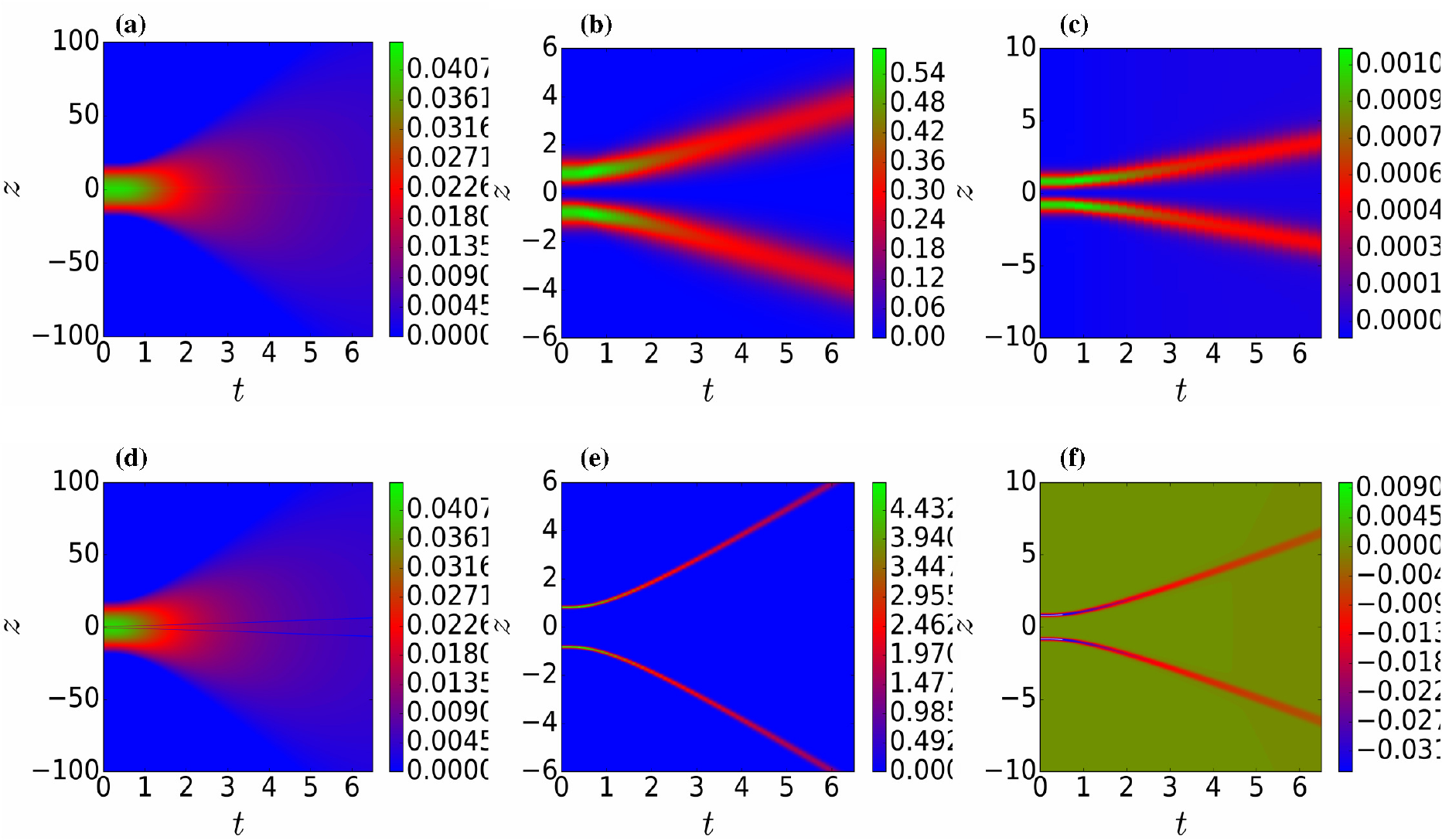} 
\protect\protect\caption{Time-of-flight evolution of BEC density (a,d), excited-impurity
density (b,e) and BEC depleted density 
$n_{\text{B}}\left(z,t\right){}_{\text{DD}}=n_{\text{B}}\left(z,t\right){}_{g_{\text{IB}}}-n_{\text{B}}\left(z,t\right){}_{g_{\text{IB}}=0}$
(c,f) versus time and position for different values of $g_{\text{IB}}=-40$
(a-c), and $g_{\text{IB}}=80$ (d-f) by using the two coupled dimensionless
Eqs.~(\ref{Eq03}) and (\ref{Eq4}) in dimensionless units. \label{Fig2} }
\end{center}
\end{figure*}

here, the first equation~(\ref{Eq03}) describes the dynamics of
the BEC, and the second equation illustrates the dynamics of the SESI.
In the above equations, $\tilde{\alpha}=l_{\text{Iz}}/l_{\text{z}}$
has the value 0.808, here $\tilde{G}_{\text{B}}=2N_{\text{B}}\omega_{\text{r}}a_{\text{B}}/\omega_{\text{z}}l_{\text{z}}$,
and $\tilde{g}_{\text{IB}}=2a_{\text{IB}}\omega_{\text{r}}f\left(\omega_{\text{Ir}}/\omega_{\text{r}}\right)/\omega_{\text{z}}l_{\text{z}}$
are the dimensionless Rb-Rb and Rb-Cs coupling strengths, respectively.
By using above mentioned experimental values, we obtained the dimensionless
Rb-Rb and Rb-Cs coupling strengths as $\tilde{G}_{\text{B}}=4.71$
and $\tilde{g}_{\text{IB}}=0.16$, respectively. From here on, we
will drop all the tildes for simplicity. To find the numerical excited state of a Cs impurity, we start with a trial  
excited state wave function for the impurity as summarized in appendix ~\ref{A1}, here 
the impurity dimensionless energy $E_{\text{I}}$ depends upon the dimensionless
imaginary time.

In order to determine the equilibrium excited-impurity imprint on
the condensate wave function, we solve numerically the two coupled
dimensionless quasi 1DGPE~(\ref{Eq03}) for the BEC and the differential
equation~(\ref{Eq4}) for the excited-impurity by using the split-operator
method \cite{Vudragovic12,Kumar15,Loncar15,Sataric16}. In this way, we demonstrate
that the $^{133}\text{Cs}$ excited-impurity leads to two bumps or
two holes at the center of the $^{87}\text{Rb}$ BEC density for attractive
or repulsive interspecies coupling strength $g_{\text{IB}}$ as displayed
in Fig.~\ref{Fig1}(a) and Fig.~\ref{Fig1}(b), respectively. For
stronger attractive $g_{\text{IB}}$ values two bumps can increase
further as depicted in Fig.~\ref{Fig1}(a), but for strong repulsive
$g_{\text{IB}}$ values two dips in the BEC density gets deeper and
deeper until BEC fragmented into three parts as illustrated in Fig.~\ref{Fig1}(b). 
Additionally, the SESI effective mass increases quadratically for
interspecies coupling strength $g_{\text{IB}}$ as presented in appendix ~\ref{A2}.
In this manuscript, we have utilized the zero-temperature GP mean-field
theory, however, as a matter-of-fact, elementary excitations can arise
from the thermal and/or quantum fluctuations \cite{PhysRevB.53.9341},
and the BEC dynamics may be considerably affected by the motion of
the excited atoms around it (thermal cloud), and by the dynamical
BEC depletion \cite{PhysRevLett.79.3553}. To give a rough estimate
to our reader, first of all, we let that the excited-impurity in our
proposed model does not affect the mean-field description of our system.
The mean-field approximations hold so long as the impurity-BEC interaction does not significantly deplete the condensate,
leading to the condition \cite{PhysRevA.70.013608,Bruderer08,PhysRevA.89.053617} 
\begin{equation}
 |a_{\text{IB}}| \xi^{-1}\ll 1 .\label{Eq4a}
\end{equation}
Here, $\xi^{-1}= l_{\text{r}}/\sqrt{2 n_{\text{1D}} a_{\text{B}} }$ is the 1D healing length. 
The dimensionless peak density of the BEC at the center of the condensate is 
$\tilde{n}_{\text{1D}}=0.355 (0.164)$ for the dimensionless Rb-Rb coupling strength 
$\tilde{G}_{\text{B}}=10 (100)$ and the corresponding value
$|a_{\text{IB}}| \xi^{-1}= 0.0020(0.0014)$, respectively. Which shows that our treatment of the single excited-impurity in a BEC system 
neglects the phenomenology of strong-coupling physics, e.g., near a Feshbach
resonance \cite{PhysRevA.88.053632}, which lies beyond 
the parameter range of Eq.~(\ref{Eq4a}). Therefore, we restrict the following calculation of the validity range
of the mean-field analysis to a BEC without any excited-impurity.
Additionally, in appendix ~\ref{A3}, we regulate how quantum and thermal fluctuations
within the Bogoliubov theory restrict the validity range of our mean-field
description.  

\begin{figure*}
\begin{center}

\includegraphics[width=14cm,height=7.5cm]{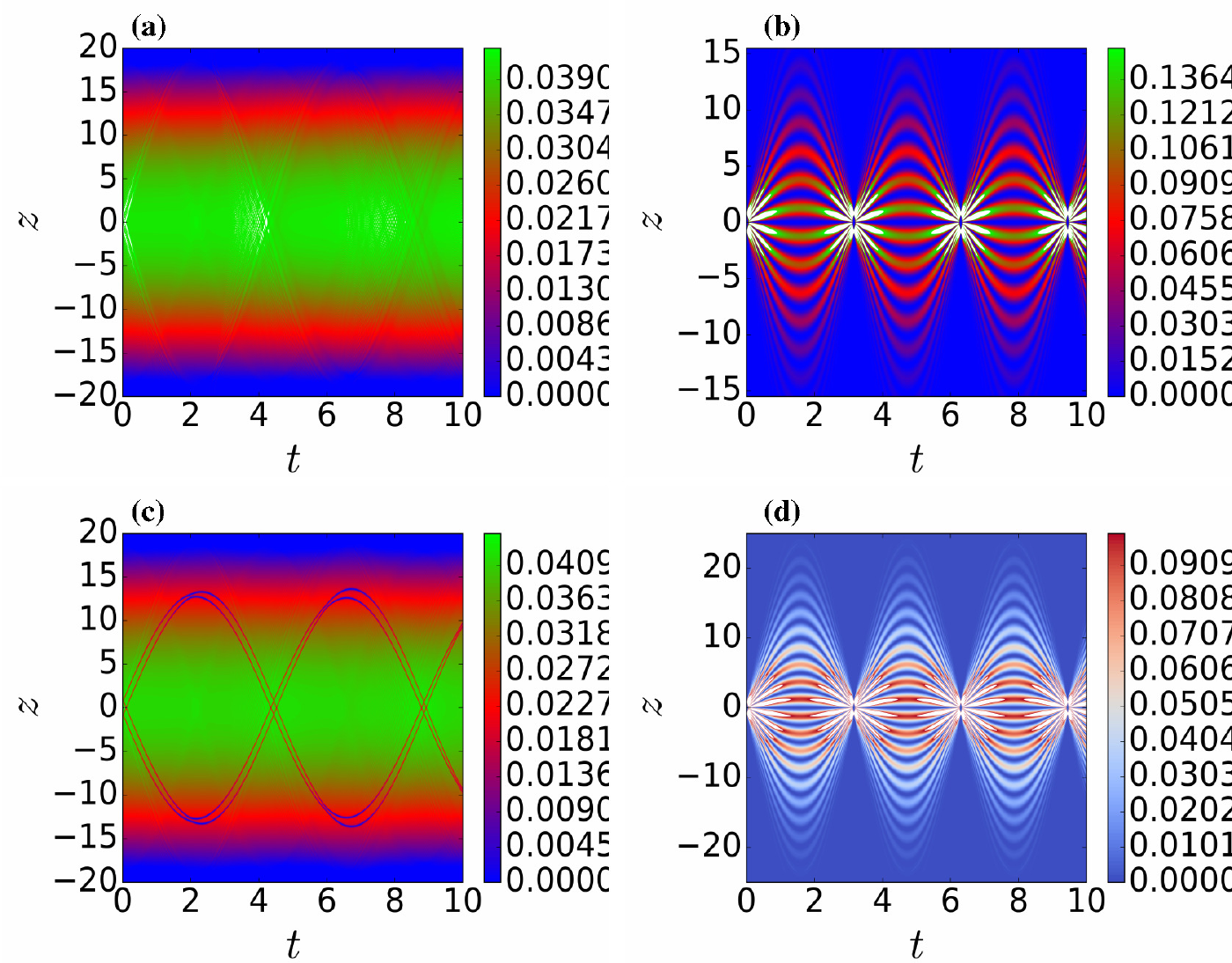}
\end{center}
\caption{Evolution of the BEC density (a,c) and the excited-impurity
density (b,d) in a harmonic trap, when the interspecies interaction
strength is switched off, versus time and position for different values
of $g_{\text{IB}}=-40$ (a-b) and $g_{\text{IB}}=80$ (c-d) in dimensionless
units. \label{Fig3} }
\end{figure*}

\section{Dynamics of the BEC and the excited-impurity}

\label{PS3}To investigate the dynamical evolution of the condensate
wave function and the excited-impurity, we investigate numerically
two quench scenarios. In the first scenario, we investigate the standard
time-of-flight (TOF) expansion after having switched off the external
harmonic trap when the excited-impurity and BEC interspecies interaction
strength is still present. In the second case, we consider an inverted
situation where the excited-impurity and the BEC interspecies interaction
strength is turned off by letting the harmonic confinement switched
on. This represents an interesting scheme to generate matter waves like
shock-waves or solitons depending on either the initial excited-impurity
and BEC interaction strength is attractive or repulsive.

In the first scheme, we turn off the magnetic trap at time $t=0$,
the BEC and the SESI is allowed to expand in all directions. At $t=0$,
the confining potential vanishes, and further acceleration results
from inter- and intra-species interactions strength. For the attractive or
repulsive interspecies coupling strength, two bumps or dips decay
slowly during the temporal evolution as shown in Figs.~\ref{Fig2}(a)
and \ref{Fig2}(d). The relative speed of decaying of these bumps
or dips from each other is zero. The SESI imprint bumps or dips are
not only decaying but also moving away from their stationary positions
as demonstrated in Figs.~\ref{Fig2}(c) and \ref{Fig2}(f).

\begin{figure*}
\begin{center}
\includegraphics[width=16cm,height=7cm]{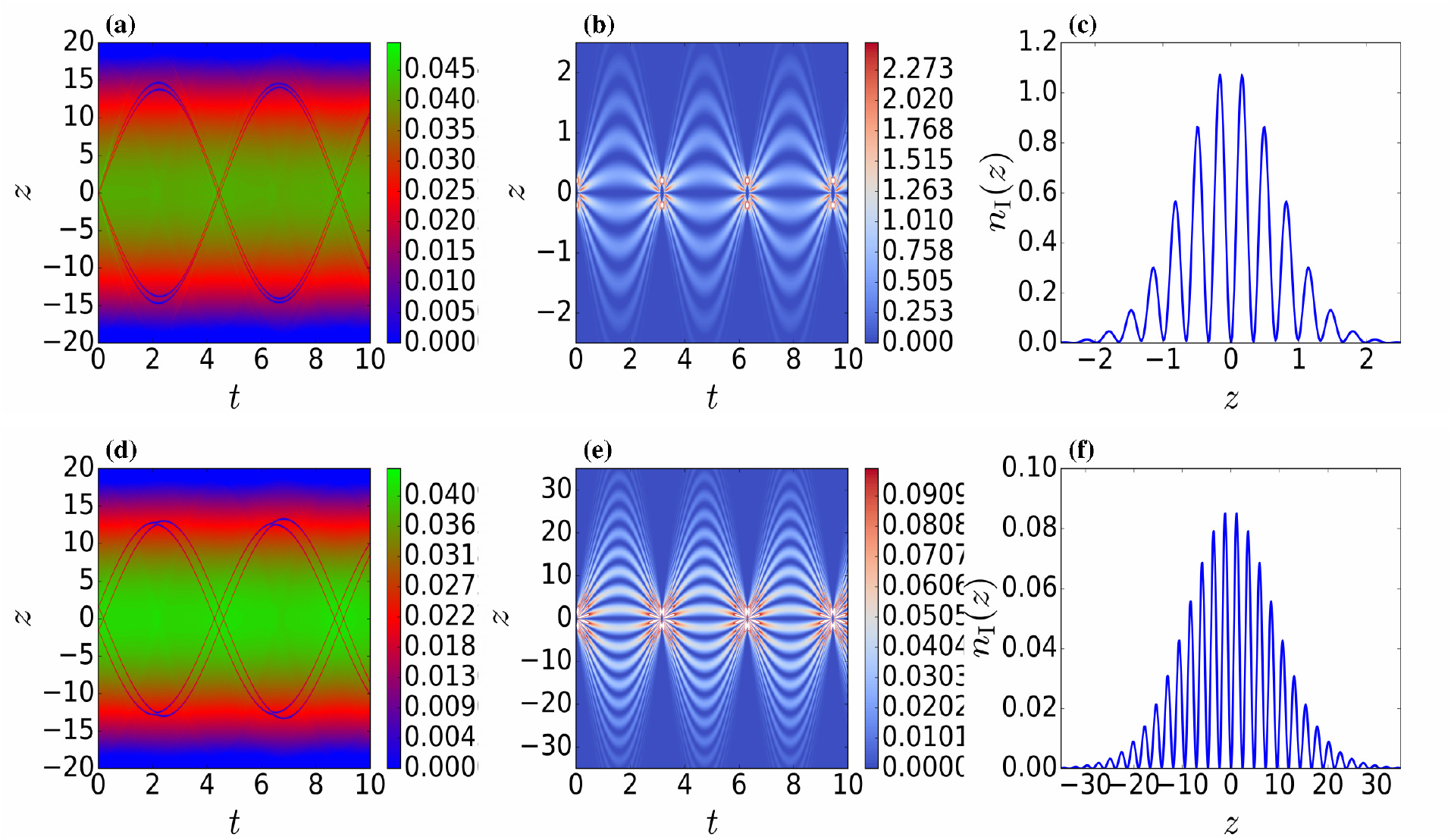}
\caption{Dynamical evolution of the BEC density (a,d) and the
excited-impurity density (b,e), when interspecies interaction strength
is switched off, versus time and position for values of $\alpha=0.2$
(a-c) and $\alpha=1.5$(d-f) in dimensionless units. In figure (c)
and (f) we plot the excited-impurity density at dimensionless time
$t=10$. \label{Fig4} }
\end{center}
\end{figure*}

In the second scenario, we introduce a numerical model of matter-wave
self-interference resulting from the attractive and repulsive interspecies
strength is switched off and within a remaining harmonic confinement,
which leads to the shock waves and gray quad-solitons, respectively,
as predicted in Figs.~\ref{Fig3}(a) and \ref{Fig3}(c). We observed
that in every scenario, approximately $t<0.045$ dimensionless time is
required to generate shock-waves or quad-solitons as shown in Figs.~\ref{Fig3}(a)
and \ref{Fig3}(c), respectively. For an initial attractive interspecies
coupling strength $g_{\text{IB}}=-40$, we examine that two excitations
of the condensate are generated at the SESI position, which travel
in different direction with identical center-of-mass speed, are reflected
at the harmonic confinement boundaries and then collide at the SESI
position as depicted in Fig.~\ref{Fig3}(a). We have done different
calculations, by changing the value of $g_{\text{IB}}<0$, in all
cases, we observed that the appearance of shock wave structures as
shown in Fig.~\ref{Fig3}(a). The density of depleted atoms around shocks becomes, at most, larger
than depletion density far away from perturbations. The corresponding
excited-impurity self-interference pattern starts breathing with dimensionless frequency $\omega_{I}/\omega_{{\rm z}}=2$
at the center of the harmonic trap as shown in Fig.~\ref{Fig3}(b). For small attractive/repulsive interspecies scattering strength 
the excited-impurity self-interference fringes show smaller strength as discussed in appendix ~\ref{A4}.
We can determine the breathing frequency of the SESI in a harmonic
trap by defining the single excited-impurity wave function 
$\psi_{\text{I}}\left(z,t\right)=\sqrt{2/\left[\sqrt{\pi}A(t)^{3}\right]}\, ze^{-\frac{z^{2}}{2A(t)^{2}}-iz^{2}R(t)},$
here $A(t)$ defines the dimensionless width of the SESI and $R(t)=-A'(t)/2\alpha^{2}A(t)$
describes the variational parameter which defines the momentum of the
SESI. We write the equation of motion for the width of the excited-impurity
by determining the Euler-Lagrangian equation of the system in 
\begin{align}
&& A''(t)-\frac{\alpha^{4}}{A(t)^{3}}+A(t)=0\,.\label{Eq11}
\end{align}
where the equilibrium state is $A\left(t=0\right)=A_{0}=\alpha$ and
$\alpha=l_{\text{Iz}}/l_{\text{z}}$ has the equilibrium value 0.808.
We solve Eq.~(\ref{Eq11}) and get the time dependent width of the
excited-impurity $A(t)=\sqrt{\left[\alpha^{4}+\left(A_{0}^{4}-\alpha^{4}\right)\cos(2t)+A_{0}^{4}\right]/2A_{0}^{2}}$.
Thus, in order to get the excited-impurity out of equilibrium we can let $A_{0}=\alpha\pm\delta$,
where $\delta$ is a small quantity. From the time dependent width
of the excited-impurity, we identify the dimensionless breathing oscillation
frequency to be $\omega_{I}/\omega_{{\rm z}}=2$.

And for the repulsive interspecies coupling strength $g_{\text{IB}}=80$,
we inspect gray quad-solitons, traveling with the same speed as shown
in Fig.~\ref{Fig3}(c). In the case of harmonic confinement with
a dimensionless potential $V_{\textrm{ext}}=z^{2}/2$, the frequency
of the oscillating soliton differs from the trap frequency by a factor
$\omega_{s}/\omega_{{\rm z}}=1/\sqrt{2}$ as exhibited in Fig.~\ref{Fig3}(c),
which was predicted in \cite{Reinhardt97,PhysRevLett.84.2298,Javed15,Javed15-2}
and experimentally observed in \cite{Strecker02,PhysRevLett.89.200404}. At the maxima of excited-impurity wave function two gray-solitons
are generated which have zero relative phase, and they travel in
opposite directions as displayed in Fig.~\ref{Fig3}(c). On the other
hand two gray-solitons are generated at the minima of the excited-impurity
wave function, they also have zero relative phase from each other
and zero phase difference as compared to the two gray-solitons which
were created at the maxima of excited-impurity wave function. If two
gray-solitons have a relative phase difference of zero, they attract
to each other when they come near to each other \cite{Aitchison91},
as demonstrate in Fig.~\ref{Fig3}(c) near to dimensionless time
$t\approx2.2$. This phenomena happens when two solitons reaches at
the trap boundary approximately at the same time $t\approx2.2$, the
latter one tries to cross the first one, therefore they collide with
one another and then reflected, and surprisingly this attractive phenomena
does not affect the oscillation frequency of the solitons in a harmonic
trap as depicted in Fig.~\ref{Fig3}(c). Furthermore, quad-solitons
collide at their originated position and go through each other without
any disturbance as exhibited in Fig.~\ref{Fig3}(c) near to dimensionless
time $t\approx4.2$. On the contrary, the single excited-impurity
exhibit self-interference fringes as demonstrated in Fig.~\ref{Fig3}(d).
As excited-impurity wave function has one maxima and one minima, therefore
when the repulsive interspecies coupling strength is switched off,
then the single excited state impurity self-interference patterns
are generated as demonstrated in Fig.~\ref{Fig3}(d). These self-interference
patterns represent a clear evidence for the spatial coherence of
excited-impurity, while the special patterns repeat themselves with
a unique frequency $\omega_{I}/\omega_{{\rm z}}=2$ as shown in Fig.~\ref{Fig3}(d),
which we calculated by solving Eq.~(\ref{Eq11}). We also find out
that the excited-impurity density self interference patterns does
not pass through each other at $z=0$, which is quite clear as they
do not exhibit solitonic behavior as demonstrated in Figs.~\ref{Fig3}(b)
and \ref{Fig3}(d). 

To illustrate a more general case for the generation of the solitons
by pinning the excited-impurity in the center of the BEC, we investigate
different masses impurities, which basically change nothing in our
proposed model but the value of parameter $\alpha=\sqrt{m_{\text{I}}/m_{\text{B}}}$.
For the value of $\alpha=0.2$, we witness a new kind of phenomena
where only two gray-solitons are generated as the distance between
the maxima and the minima of the excited-impurity wave function is
quite small therefore solitons get the overall shape of the sculpted
BEC as depicted in Fig.~\ref{Fig4}(a). The shape of these gray bi-solitons
is totally different than the shape of the solitons for parameter
$\alpha=0.808$ as demonstrated in Fig.~\ref{Fig3}(c). For the case
$\alpha=0.2$, near to the trap boundary, soliton does not attract each other, 
as they express unite identity, as demonstrated in Fig.~\ref{Fig4}(a). Additionally,
these bi-solitons oscillate in a harmonic trap with the same dimensionless
frequency $\omega_{s}/\omega_{{\rm z}}=1/\sqrt{2}$ as predicted in the 
previous case. In Fig.~\ref{Fig4}(d), we use $\alpha=1.5$, in this
scenario, again quad gray-solitons are generated, which reveal similar properties
as discussed for the case of $\alpha=0.808$. Furthermore, the
SESI interference fringes height decreases with increasing bare mass
of the excited-impurity and vise versa as demonstrated in Fig.~\ref{Fig4}(c)
and Fig.~\ref{Fig4}(f). On the other hand, we observe that the number of interference
fringes increases with increasing the bare mass of the excited-impurity
as illustrated Fig.~\ref{Fig4}(c) and Fig.~\ref{Fig4}(f). Additionally, as
the bare mass of the SESI increases, the width of center-of-mass of the solitons
decreases as shown in Fig.~\ref{Fig4}(a) and Fig.~\ref{Fig4}(d), the reason for this is quite simple and clear as 
solitons depth increases with increasing mass, therefore, they can not move away from their central position.

\begin{figure}
\begin{center}
\includegraphics[width=8.5cm,height=6.0cm]{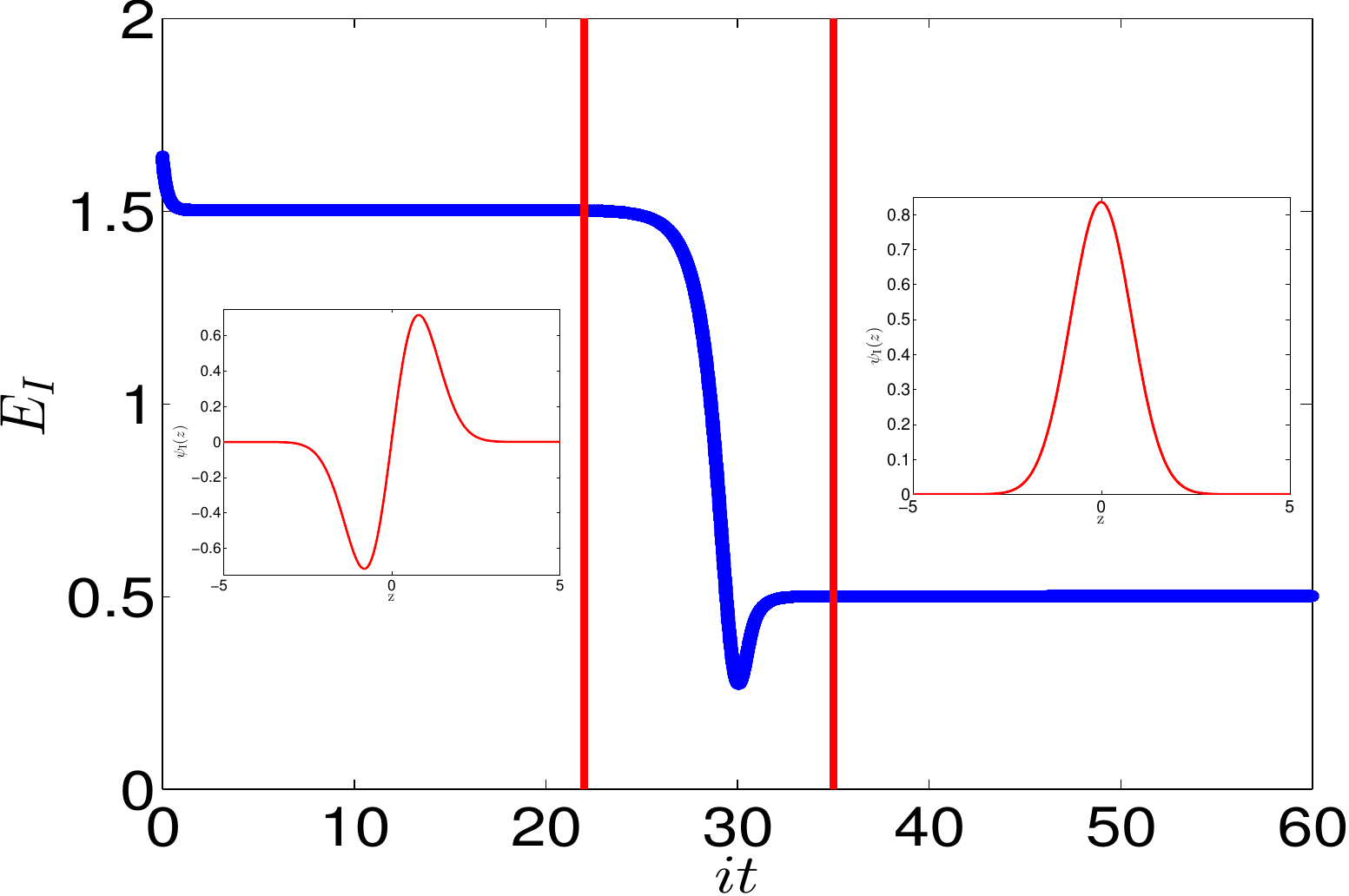} 
\end{center}
\caption{ Dimensionless energy decay in dimensionless imaginary time. Here a rectangular widow 
is separating the transition region from excited to ground state from left to
right, respectively. In inlet we plot the excited-impurity and the ground-state wave functions.  \label{Fig5} }
\end{figure}
\section{Discussion}
\label{PS6}Our studies elucidate the role of single excited state
impurity pinned in a $^{87}\text{Rb}$ BEC. We model our system in
the mean-field regime by writing the one-dimensional two coupled differential
equations, this approximation is valid for relatively weak interspecies
interaction and for single excited-impurity. We pinned the excited-impurity
in the condensate center and diminished its decay by using the quantum
zeno effect. We found out that the BEC depletion induced by the single
excited state impurity, cause the BEC density to split into three
parts. During our calculation, we have found out that the excited-impurity
imprint decays marginally for the attractive interspecies coupling
strength, and in repulsive interspecies coupling strength, it starts
decay significantly as compared to the small value of the interspecies
interaction strength. We have used the numerical simulation to analyze
generation and the dynamics of gray quad-solitons or bi-solitons in
the Bose-Einstein condensate. We disclose that the shape of newly
generated solitons depends on the bare mass of the excited state
impurity. We would like to remark that even though in our analysis
we use an idealized potential for the excited-impurity, but such an
approximation is known to, not only, encapsulate the basic physics,
but can also be a good approximation to experimental setups.

%

\appendix

\section{Impurity energy}

\label{A1}
With
$\psi_{\text{I}}\left(z,t\right)=\psi_{\text{I}}\left(z\right)e^{-i E_{\text{I}}t}$, where $E_{\text{I}}$
is the impurity dimensionless energy, we plotted in Fig.~\ref{Fig5}, the impurity dimensionless energy
vs the dimensionless imaginary time when there
is no condensate present at the background of the impurity. 
To find the numerical excited state of a Cs impurity, we start with a trial  
excited state wave function for the impurity 
\begin{align}
&& \psi_{\text{I}}\left(z\right)=\sqrt{\frac{2}{\sqrt{\pi}A^{3}}}ze^{-\frac{z^{2}}{2A^{2}}}\,,
\end{align}
here $A$ is the dimensionless width of the first excited state of the wave-packet, for the excited state $A=\alpha$
where $\alpha=l_{\text{Iz}}/l_{\text{z}}$ has the value 0.808. For the single excited state, 
the excited-impurity state is durable for $it\simeq70$. 
It means that in our numerical simulation the excited-impurity can be seen 
for a specific dimensionless imaginary time interval which later decays to 
its ground state as shown in Fig.~\ref{Fig5}.

\begin{figure}
\begin{center}
\includegraphics[width=8.5cm,height=6.0cm]{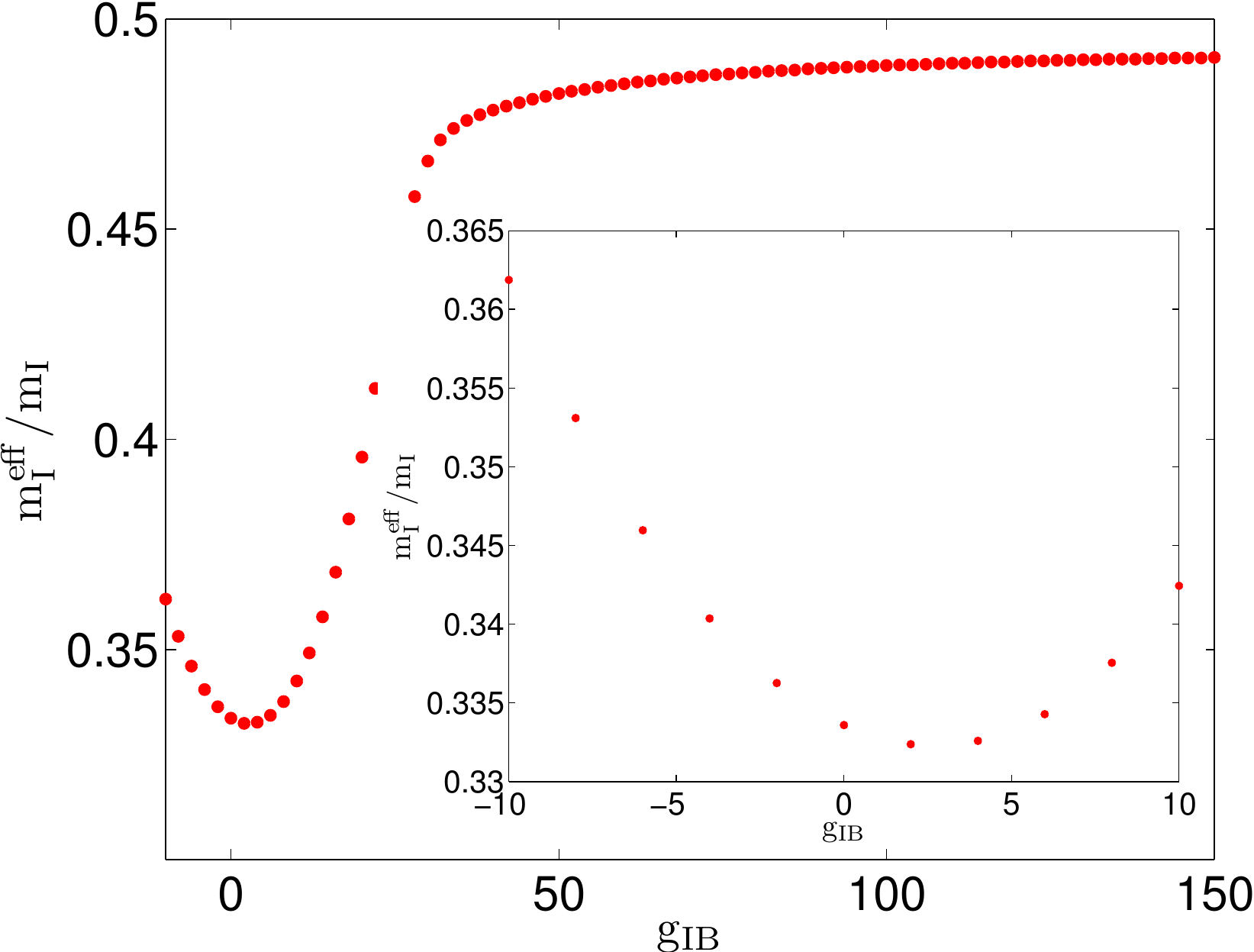}
\end{center}
\caption{(Color online) Effective mass of $^{133}\text{Cs}$ impurity versus
the impurity-BEC coupling strength $g_{\text{IB}}$. Inlet shows that
effective mass increases quadratically for small impurity-BEC coupling
strength $g_{\text{IB}}$  in dimensionless units.\label{Fig6} }
\end{figure}

\begin{figure*}
\begin{center}
\includegraphics[width=12cm,height=7.5cm]{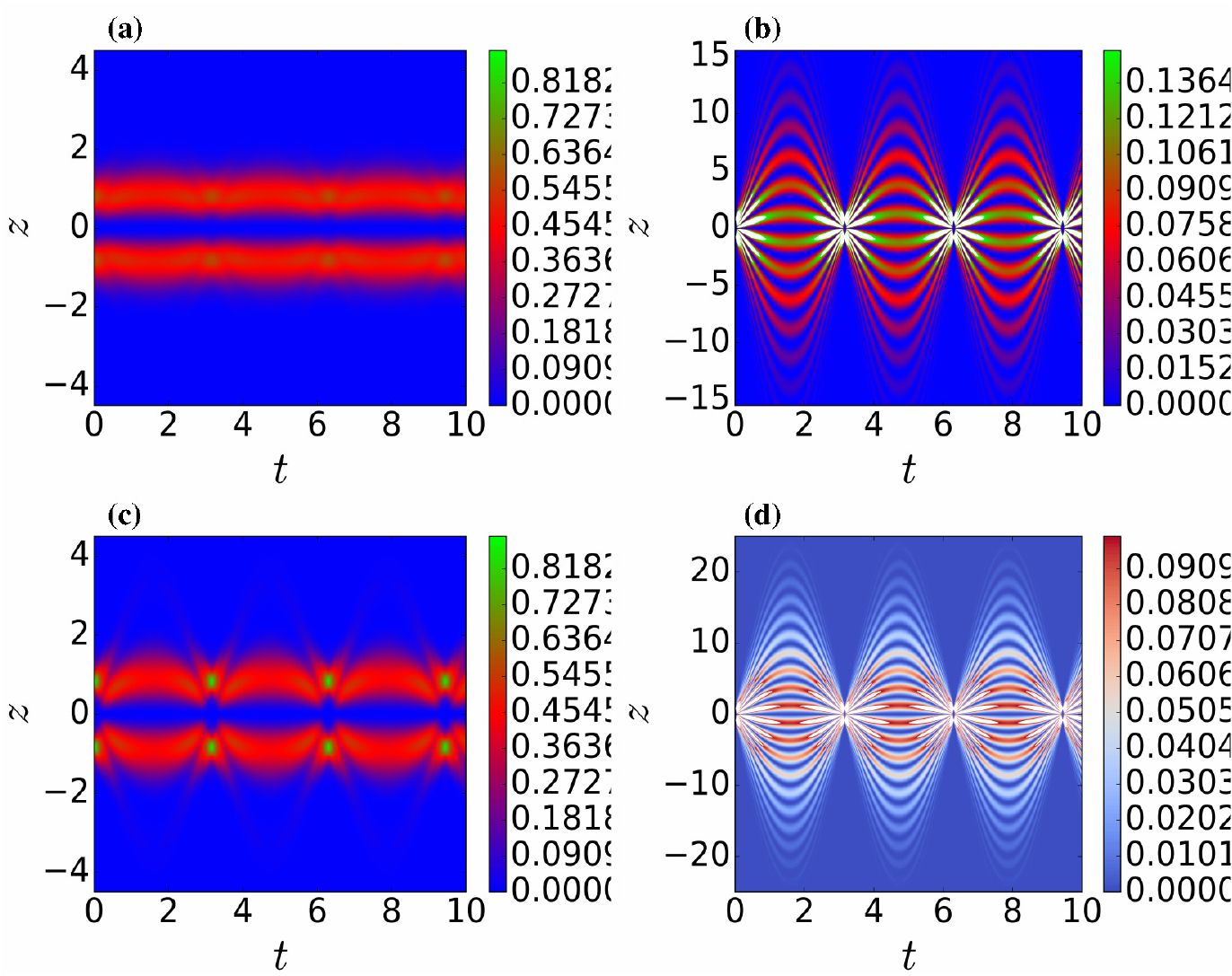}
\end{center}
\caption{Evolution of the excited-impurity
density in a harmonic trap, when the interspecies interaction
strength is switched off, versus time and position for different values
of interspecies interaction strength (a) $g_{\text{IB}}=-10$, (b) $g_{\text{IB}}=-40$,  
(c) $g_{\text{IB}}=20$ and (d) $g_{\text{IB}}=80$ in dimensionless
units. \label{Fig7} }
\end{figure*}
\section{Effective mass}
\label{A2} The effective mass of the excited-impurity is denoted as
$m_{\text{I}}^{\text{eff}}=\hbar/\left(l_{\text{Iz}}^{2}\omega_{\text{z}}\right)$,
where the excited-impurity oscillator length $l_{\text{Iz}}=\sqrt{2}\sigma$
come from the standard deviation $\sigma=\sqrt{<z^{2}>-<z>^{2}}$,
with $<\bullet>=\int\bullet\left|\psi_{\text{I}}(z)\right|^{2}dz$
representing the expectation value. Figure \ref{Fig6} shows the ratio
of the effective mass of the $^{133}\text{Cs}$ impurity with respect
to the bare mass $m_{\text{I}}$, which increases quadratically for
interspecies coupling strength $-10<g_{\text{IB}}<10$ as shown in the
inlet of Fig.~\ref{Fig6}, and becomes marginally saturated for interspecies
coupling strength $g_{\text{IB}}>80$. Here, we utilize the mean-field regime to determine the effective mass of the excited-impurity. 
Through this connection, one may extend this work to investigate polaron physics, 
in order to include the impact of quantum and thermal fluctuations \cite{PhysRevB.80.184504,PhysRevA.84.063612,Santamore11,Grusdt2015}.

\section{Mean-field analysis}
\label{A3} To give a rough estimate
to our reader, first of all we let that the excited-impurity in our
proposed model does not affect the mean-field description of our system.
Therefore, we restrict the following calculation of the validity range
of the mean-field analysis to a BEC without any excited-impurity.
In the following, we regulate how quantum and thermal fluctuations
within the Bogoliubov theory restrict the validity range of our mean-field
description.

\subsection*{Quantum depletion}
According to the Bogoliubov theory the three-dimensional quantum fluctuation
term is defined in Thomas-Fermi approximation: 
\begin{align}
 && n_{\text{QF}}^{\text{3D}}\left(\mathbf{r}\right)=n_{\text{0B}}^{\text{3D}}\left(\mathbf{r}\right)
 -n_{\text{B}}^{\text{3D}}\left(\mathbf{r}\right)=\frac{8}{3\sqrt{\pi}}\left[N_{\text{B}}a_{\text{B}}n_{\text{B}}
 ^{\text{3D}}\left(\mathbf{r}\right)\right]^{3/2}.\label{eq3}
\end{align}
We assume an effective one-dimensional setting with $\omega_{\text{z}}\ll\omega_{\text{r}}$,
so we decompose the BEC wave-function $\psi_{\text{B}}(\mathbf{r},t)=\psi_{\text{B}}(z,t)\phi_{\text{B}}(\textbf{ r}_{\perp},t)$
with $\textbf{ r}_{\perp}=\left(x,\; y\right)$ and %
\begin{align}
&& \phi_{\text{B}}(\textbf{ r}_{\perp},t) =\frac{e^{-\frac{x^{2}+y^{2}}{2l_{\text{r}}^{2}}}}{\sqrt{\pi}l_{\text{r}}}
e^{-i\omega_{\text{r}}t}\,.\label{eq2}
\end{align}
We integrate out the transversal degrees of freedom from equation
(\ref{eq3}) to get an effective one-dimensional setting 
\begin{align}
 && n_{\text{QF}}^{\text{1D}}\left(z\right)=n_{\text{0B}}^{\text{1D}}\left(z\right)-n_{\text{B}}^{\text{1D}}\left(z\right)
 =\frac{16}{9\pi l_{\text{r}}}\left[N_{\text{B}}a_{\text{B}}n_{\text{B}}^{\text{1D}}\left(z\right)\right]^{3/2}.\label{eq3-1}
\end{align}
We know that for larger inter-particle interaction strength the BEC
density is characterized by the Thomas-Fermi (TF) profile $n_{\text{B}}^{\text{1D}}\left(z\right)=\frac{\mu^{\text{1D}}}{G_{\text{B}}^{\text{1D}}}\left(1-\frac{z^{2}}{R_{\text{z}}^{2}}\right)$
with $R_{\text{z}}^{2}=\frac{2\mu^{\text{1D}}}{m_{\text{B}}\omega_{\textrm{z}}^{2}}$.
With this we calculate the one-dimensional quantum fluctuation depleted
term with respect to the number of particles \textbf{$N_{\text{B}}=200$} by
using $N_{\text{QF}}^{\text{1D}}=\int n_{\text{QF}}^{\text{1D}}\left(z\right)dz$
and get 
\begin{align}
 && \frac{N_{\text{QF}}^{\text{1D}}}{N_{\text{B}}}=\frac{3^{1/3}}{4}
 \left(\frac{a_{\text{B}}^{4}N_{\text{B}}}{l_{\text{r}}^{2}l_{\text{z}}^{2}}\right)^{1/3}.\label{eq4}
\end{align}
We evaluate this relative depletion for the system parameters of our
study. With this we obtain from (\ref{eq4}) $\frac{N_{\text{QF}}^{\text{1D}}}{N_{\text{B}}}=0.0022$,
so that the quantum fluctuations are, indeed, negligible. 

\subsection*{Thermal depletion}

Correspondingly, the one-dimensional thermal depleted term with respect
to the number of particles follows from Bogoliubov theory to be 
\begin{align}
 && \frac{N_{\text{TF}}^{\text{1D}}}{N_{\text{B}}}=\gamma\left(\frac{T}{T_{\textrm{c}}}\right)^{2}\label{eq:13}
\end{align}
with the dimensionless prefactor 
\begin{align}
 && \gamma=\frac{5^{2/5}\pi^{2}}{2^{3/2}\times3^{3/5}}\left\{
 \frac{N_{\text{B}}^{1/3}}{\xi\left(3\right)^{8/3}}\frac{a_{\text{B}}^{2}}{l_{\text{r}}
 ^{4/3}l_{\text{z}}^{2/3}}\right\} ^{1/5}.\label{eq12-1-1}
\end{align}
For our system parameters we obtain $\gamma=0.046$ and the critical
temperature $T_{\textrm{c}}=\frac{\hbar}{k_{\text{B}}}\left(\omega_{r}^{2}\omega_{\textrm{z}}\frac{N_{\text{B}}}
{\xi\left(3\right)}\right)^{1/3}=14.7~\text{nK}$.
Thus, choosing a reasonable ratio of the thermal depleted term $\frac{N_{\text{TF}}^{\text{1D}}}{N_{\text{B}}}=0.001$,
we estimate the temperature of the system to be 
\begin{align}
 && T=T_{\textrm{c}}\sqrt{\frac{1}{\eta}\frac{N_{\text{TF}}^{\text{1D}}}{N_{\text{B}}}}=2.15~\text{nK}\label{eq12-1-1-1}
\end{align}
With this we conclude that, if the temperature of the system is lower
than $T=2.15~\text{nK}$, the thermal fluctuations are not affecting
the Bose-Einstein condensate.

\section{Excited-impurity self-interference patterns}
\label{A4} The single excited-impurity wave packet display self-interference
patterns as demonstrated in Fig.~\ref{Fig7}.
As excited-impurity wave function has one maxima and one minima, therefore
when the attractive/repulsive interspecies coupling strengths are switched off,
then the excited-impurity self-interference patterns
are generated. We find out that the excited-impurity density self interference patterns does
not pass through each other at $z=0$, which is quite clear as they
do not exhibit solitonic behavior as demonstrated in Fig.~\ref{Fig7}. 
As we can see from the Fig.~\ref{Fig7}, for small attractive/repulsive interspecies scattering strength 
the excited-impurity self-interference fringes demonstrate smaller strength and vise versa.

\section{Acknowledgment}

We gratefully acknowledge support from the \textcolor{red}{German
Academic Exchange Service (DAAD)}. We thank Thomas Busch and Axel
Pelster for insightful comments.

\bibliographystyle{apsrev4-1}
\bibliography{ExImpurity.bib}

\end{document}